# COVID-19 Mobility Data Collection of Seoul, South Korea

## Authors
Jungwoo Cho[1,2], Soohwan Oh[1,2], Seyun Kim[1], Namwoo Kim[1], Yuyol Shin[1], Haechan Cho[1], Yoonjin Yoon[1*]

## Affiliations
[1] Department of Civil and Environmental Engineering, Korea Advanced Institute of Science and Technology, 34141, Republic of Korea.
[2] These authors contributed equally: Jungwoo Cho, Soohwan Oh
[*] corresponding author(s): Yoonjin Yoon (yoonjin@kaist.ac.kr)

## Abstract

The relationship between pandemic and human mobility has received considerable attention from scholars, as it can provide an indication of how mobility patterns change in response to a public health crisis or whether reduced mobility contributes to preventing the spread of an infectious disease. While several studies attempted to unveil such relationship, no studies have focused on changes in human mobility at a finer scale utilizing comprehensive, high-resolution data. To address the complex association between pandemic's spread and human mobility, this paper presents two categories of mobility datasets—trip mode and trip purpose—that concern nearly 10 million citizens' movements during COVID-19 in the capital city of South Korea, Seoul, where no lockdown has been imposed. We curate hourly data of subway ridership, traffic volume and population present count at selected points of interests. The results to be derived from the presented datasets can be used as an important reference for public health decision making in the post COVID-19 era.

## Background & Summary

Investigating the relationship between epidemic outbreak and human mobility has received significant attention from researchers around the world[1-4]. Using province-to-province travel history data[3-4], Kraemer et al. (2020) and Tian et al. (2020) examined the effect of human mobility and control measures on the COVID-19 epidemic in China[1,2]. While the studies provided an important indication of how reduced human mobility contributed to mitigating the spread of COVID-19, no study focused on the relationship between COVID-19 and mobility at a finer scale using comprehensive, high-resolution datasets.

Seoul, the capital city of South Korea, has been able to maintain the maximum daily case under 52 (5.34 per 1M population) as of July 26, 2020 without a major lockdown or restriction on movement[5-8]. During the COVID-19 outbreak, the way that 10 million citizens travel has undergone considerable changes in accordance with increased public health risks across the nation, exhibiting an unprecedented drop in people's movement even without intensive mobility-constraining measures[9-11].

To address the complex association between pandemic's spread and mobility pattern changes in a highly populated city, this paper presents a comprehensive set of mobility data concerning nearly 10 million citizens' movements during COVID-19 in Seoul. Specifically, we collect and curate two categories of mobility datasets: trip mode and trip purpose. The trip mode datasets contain hourly subway ridership[12] at 275 stations and hourly traffic volume[13] at 105 traffic count locations across Seoul. Each dataset is a representative of individual movements within the city, accounting for 7.47 million daily trips on average for the case of subway ridership. The trip purpose dataset contains the number of people who are physically present in a specific geospatial extent during each one-hour period (i.e. *population present count*[14]). Population present count data is collected and estimated by Korea Telecom (KT)[15] using the company's geolocation data of people. The data is available at the smallest statistical unit[*] (SSU) level, for which population counts are available on both resident- and present-basis. This dataset can be used as a proxy for capturing mobility patterns at specific points of interests (POIs). A POI is associated with categories of places, such as restaurants and shopping malls[16,17], that people visit for particular purposes. In this paper, we select several POI categories and retrieve population present count data of SSUs in which the selected POIs are located. The retrieved data can be used to examine how human mobility within small regions containing POIs have changed in association with increased public health risks.

Trip mode dataset spans a 185 day-duration for both year 2020 and year 2019 while trip purpose dataset spans a 100 day-duration. One can use last year's data as a baseline to analyze the reduction in mobility while accounting for seasonality. By analyzing mobility patterns within the city that has not imposed any major lockdown or movement restrictions, one can form an important basis for understanding the relationship between COVID-19 spread and changes in daily mobility patterns, which can further be used as a reference for public health decision making in the post COVID-19 era.

---

[*] There are 19,153 SSUs (i.e. *jipgyegu*) in Seoul. The area of SSU ranges from 301 $m^2$ to 9,691,745 $m^2$ (mountainous region) with a median of 11,689 $m^2$.

## Methods

The mobility dataset is categorized into two types: trip mode and trip purpose. The following subsections provide detailed descriptions of data collection and curation. The datasets are collected from Seoul Open Data Plaza, Seoul Metro and Seoul Transport Operation & Information Service (TOPIS), all of which offer free copyright licenses for the creation of secondary works[12,13,15]. Summaries of data descriptions and data sources are shown in Table 1 and Table 2.

### *Trip mode data*

The trip mode dataset comprises subway ridership and traffic volume data measured at hourly intervals. Subway ridership data, provided as an open dataset by Seoul Metro[12], consists of the hourly number of passengers getting on and off at 275 stations. The modal share by the subway system is 39.9% in Seoul[18], accounting for 7.47 million daily trips on average in 2019[12]. Traffic volume data consists of hourly traffic volumes at 100 traffic count locations, such as major road links, including downtown, city border, bridge and city highway, and is measured and provided by Seoul TOPIS[13].

In this paper, we provide subway ridership and traffic volume data for both year 2020 and 2019, so that one can use last year's data as a baseline to analyze the reduction in mobility while accounting for seasonality. In each dataset, we calculate the percentage change in hourly subway ridership and traffic volume in 2020 compared to those in 2019, by taking the difference of data in 2020 and 2019 for each corresponding day of week. For example, the data on Feb 7, 2020 (Friday) corresponds to the data on Feb 8, 2019 (Friday).

### *Trip purpose data*

Trip purpose dataset includes the number of people who are physically present in a SSU in Seoul during each one-hour period, which we call *population present count*[14]. Unlike usual resident population counts, population present counts include daytime, workplace and visitor populations and hence can serve as a proxy to analyze people's movements at 1-hour intervals.

Population present count data is collected and estimated by Korea Telecom (KT) based on the company's geolocation data of people[15,19,20]. People's geolocation data is collected at 8,266 telecommunication base stations operated by KT across Seoul[19,20]. Whether a person is present within the service coverage of a base station is obtained by identifying at which base station the most recent mobile phone signal is captured[19,20]. Once the population count is measured at each base station, it is then mapped to SSUs encompassing 604 km$^2$ area of Seoul. An estimation model was developed by KT to distribute population counts to SSUs, considering additional datasets such as residential population, business working population, land use and public transportation ridership[19,20]. For the detailed description of the model, one can refer to KT's registered patent[20]. The population present count is further modified to take into account KT's mobile phone market share (32%), LTE subscription rate (89%) and the ratio of turned-on devices (93%)[19,20].

Population present count data derived from mobile phone signals can serve as a proxy for capturing changes in people's travel behavior. For understanding how mobility patterns have changed at a local level, we extract Points of Interests (POI) location data using Google Places API based platform[22]. Point of interest (POI) represents the geographical location of a particular place that people congregate for a specific purpose. In this paper, we select Starbucks stores, Michelin-listed restaurants, department stores and workplaces as POIs and retrieve population present count data for each POI categories.

Google Places API's Nearby Search Request[21] is a widely-used platform to find the coordinates of places that match a search term (i.e. Starbucks in Seoul) within a certain radius of a reference point. As Google Places API limits the number of search results to a maximum of 60, we used a Google Places API based platform, Apify[22], to obtain the location information of POIs. Specific parameters used for data crawling are (126.978, 37.5665) for viewport point latitude and longitude, zoom level 10 and unspecified maximum crawled places, given a search string containing "name of POI" (i.e. Starbucks) plus "in Seoul" in English. Using this platform, we obtained the coordinates of the POIs listed below. Once the POI coordinates are found, we conducted a geometric operation to determine the intersection between POI coordinates and geographic regions bounded by SSUs. We performed *spatial join* of POI point layer and SSU polygon layer using QGIS[23] and aggregated the population present counts of a set of SSUs where the desired POIs are located. The description of each POI category is as follows:

- **Starbucks stores**: Starbucks is the leading coffee chain in South Korea, having approximately 35% ($12.4b/$35.2b in terms of sales revenue) of the nation's coffee franchise market[24]. There are 512 Starbucks stores in Seoul, indicating 52.4 stores per million population. South Korea is known to be one of the world's top five countries with the largest number of Starbucks stores: 1,405 stores as of May 2020. Most of the stores are located in urban centers, especially those with high density population. There are 333 SSUs containing the POIs, and the area of SSUs ranges from 0.0029 $km^2$ to 0.98 $km^2$ with a median of 0.046 $km^2$.
- **Michelin restaurants**: We select 80 restaurants that are currently or were previously Michelin-starred or Bib Gourmand-listed in Seoul. The majority of these restaurants are located in regions populated with other restaurants. There are 65 SSUs containing the POIs, and the area of SSUs ranges from 0.0075 $km^2$ to 1.15 $km^2$ with a median of 0.093 $km^2$.
- **Department stores**: A total of 17 department stores in the metropolitan Seoul are selected as POI. This POI represents geolocations populated with major retail stores serving a high volume of people during the daytime. There are 16 SSUs containing the POIs, and the area of SSUs ranges from 0.021 $km^2$ to 0.85 $km^2$ with a median of 0.077 $km^2$.
- **Workplaces**: Geospatial regions concentrated with workplaces are determined based on the demographic information of travel survey data[25]. Among 424 transportation analysis zones (TAZ) in Seoul, the top 5 TAZs with the highest number of working populations were selected. These regions represent central business districts and industrial complexes in Seoul. We extract population present counts of 175 SSUs comprising the top 5 TAZs. The area of SSUs ranges from 0.0019 $km^2$ to 1.96 $km^2$ with a median of 0.019 $km^2$.

*Epidemiological data*
In addition to mobility data, we collect epidemiological data from daily briefing video clips and documents published by Seoul Metropolitan Government[26]. The epidemiological data consists of the number of cumulative confirmed cases, cumulative fatality cases, active cases and cumulative tests conducted in Seoul.

## Data Records

The mobility datasets are released as comma-separated values (CSV) files, each of which includes the intra-city mobility records from Jan 20, 2020 to Apr 28, 2020 for trip purpose dataset (July 26, 2020 for trip mode dataset) and as well as from Jan 21, 2019 to Apr 30, 2019 (July 28, 2020 for trip mode dataset). The collection of trip mode and trip purpose mobility data can be download from *figshare*[27]. A detailed description for each data category is included in Table 1. Summary descriptive statistics are shown in Table 2. A general data format is as follows:

- ***day***: elapsed day over time since the first case recorded on Jan 19, 2020
- ***date***: date of the elapsed day
- ***dayofweek***: day of week of the elapsed day
- ***cumulative confirmed***: cumulative number of confirmed cases in Seoul
- ***active cases***: number of active (i.e. quarantined) cases in Seoul
- ***cumulative fatality cases***: cumulative number of fatality cases in Seoul
- ***total tests conducted***: total number of tests conducted in Seoul
- ***hour***: hour
- ***data2019***: hourly count of subway ridership, traffic volume, and POI population count in 2019
- ***data2020***: hourly count of subway ridership, traffic volume, and POI population count in 2020
- ***percentage_change***: percentage change in hourly subway ridership, traffic volume, and POI population count in 2020 compared to those in 2019.

**Table 1.** Descriptions of epidemiological and mobility datasets

| Category | Name | Data field | Data frequency | Description | Source |
|---|---|---|---|---|---|
| Epidemiology | Epidemiology | day, date, dayofweek, cumulative confirmed, active cases, cumulative fatality cases, total tests conducted | Daily | Number of cumulative confirmed, fatality cases, active cases and total tests conducted | Seoul Metropolitan government[26] |
| Mobility (trip mode) | Subway ridership | day, date, dayofweek, hour, data2019, data2020, percentage change | Hourly | Hourly number of people getting on and off at subway stations using transportation cards | Seoul Metro[12] |
| Mobility (trip mode) | Traffic volume | day, date, dayofweek, hour, data2019, data2020, percentage change | Hourly | Hourly traffic volume at 105 inbound or outbound count locations | Seoul TOPIS[13] |
| Mobility (trip purpose) | Starbucks | day, date, dayofweek, hour, data2019, data2020, percentage change | Hourly | Population present counts at a set of SSUs containing specific POIs | Seoul Open Data Plaza[15] |
| Mobility (trip purpose) | Michelin restaurants | day, date, dayofweek, hour, data2019, data2020, percentage change | Hourly | Population present counts at a set of SSUs containing specific POIs | Seoul Open Data Plaza[15] |
| Mobility (trip purpose) | Department stores | day, date, dayofweek, hour, data2019, data2020, percentage change | Hourly | Population present counts at a set of SSUs containing specific POIs | Seoul Open Data Plaza[15] |
| Mobility (trip purpose) | Workplaces | day, date, dayofweek, hour, data2019, data2020, percentage change | Hourly | Population present counts at a set of SSUs containing specific POIs | Seoul Open Data Plaza[15] |

**Table 2.** Summary statistics for each mobility dataset

| Category | | | Min. | 1st Q. | Median | Mean | 3rd Q. | Max. | S.D. |
|---|---|---|---|---|---|---|---|---|---|
| Trip mode | Subway ridership | | 2,396,254 | 4,898,177 | 7,181,068 | 6,892,063 | 8,142,522 | 10,681,708 | 2,192,870 |
| | Traffic volume | | 2,153,180 | 3,123,365 | 3,343,283 | 3,212,041 | 3,446,938 | 3,651,934 | 359,528 |
| Trip purpose | Starbucks | Daily population count | 347,250 | 490,058 | 655,201 | 614,635 | 683,566 | 768,948 | 114,880 |
| | Michelin restaurants | Daily population count | 78,424 | 105,560 | 141,210 | 134,745 | 152,216 | 181,149 | 28,862 |
| | Department stores | Daily population count | 20,977 | 27,842 | 30,314 | 30,297 | 32,195 | 38,998 | 4,320 |
| | Workplaces | Daily population count | 162,515 | 231,267 | 377,131 | 339,805 | 397,230 | 447,379 | 88,287 |

# Technical Validation
*Missing value treatment and imputation*

In the case of traffic volume data, there can be a high proportion of missing values due to the temporary failure of equipment including loop detectors and radar guns. To treat these missing values, we excluded all data collected from a single traffic count location if missing values are more than 20% of the sample. This leaves 241 traffic count locations (inbound or outbound) out of 270. If less than 20% of the collected data are missing, we imputed missing values using K-nearest neighbourhood[28], where we set *K*=5 and used 4 variables including date, day of week, hour, and weekday or weekend to calculate the Euclidean distance between data points. The same process is also applied to population present count, where data for three census tracts were removed as more than 20% of the sample were missing values. No missing values were found in the subway ridership dataset.

# Usage Notes
There are several holiday periods during the horizon from Jan 20, 2020 to July 28, 2020 and also for the corresponding days in 2019. For example, new Year holiday periods were Jan 25-28 in 2019 and Feb 3-5 in 2020. As travel behavior changes during holidays, we provide the descriptions of holidays in a separate file so that one can either replace or ignore any data in the concerned period.

# Code Availability
The R programming language scripts used for missing value treatment are publicly available and can be freely downloaded from *figshare*.

# Acknowledgements

# Author contributions
Data curation: All authors contributed to curating the database. Technical validation: S.O., J.C. Conceptualization, Resources, Supervision: Y.Y. Writing - Original Draft: J.C., S.O. Writing - Review & Editing. Y.Y.

# Competing interests
The authors declare that they have no conflict of interest.